\documentclass[12pt]{article}
\usepackage[margin=1.0in]{geometry}
\usepackage[utf8]{inputenc}
\usepackage{graphicx}
\usepackage{float}
\usepackage{amsmath}
\usepackage{esint}
\usepackage{hyperref}
\usepackage{color}
\usepackage{natbib}

\title{Obituary, Professor John D. Barrow 1952--2020\footnote{Published in {\it Astronomy \& Geophysics}, Volume 61, Issue 6, December 2020, Pages 6.11–6.12, \url{https://doi.org/10.1093/astrogeo/ataa081}}\\The Sharpest of Minds}
\author{Jo\~ao Magueijo$^1$ and John Webb$^2$}
\date{}
\begin{document}
\maketitle
\noindent\small{$^1$Theoretical Physics Group, The Blackett Laboratory, Imperial College, Prince Consort Rd., London, SW7 2BZ, UK.}\\
\small{$^2$Clare Hall, University of Cambridge, Herschel Rd, Cambridge CB3 9AL, UK.}\\ \\

On Saturday 26 September, around 4am, John Barrow died aged 67, with his wife Elizabeth and son Roger at his side. From a scientific perspective, it is hard to conceive a more premature end. During lockdown alone, whilst undergoing chemotherapy and in the full knowledge that his cancer was inoperable, John managed to co-author 11 scientific papers and write a new book (``One Plus One''). Even by his own standards of productivity, this is staggeringly impressive, an achievement he was openly proud of. From a broader perspective, with a wife, 3 children and 5 young grandchildren, many strong friendships, and so much more to offer the world, he departed far too soon. \\

\begin{figure*}
    \centering
    \includegraphics[width=1.0\textwidth]{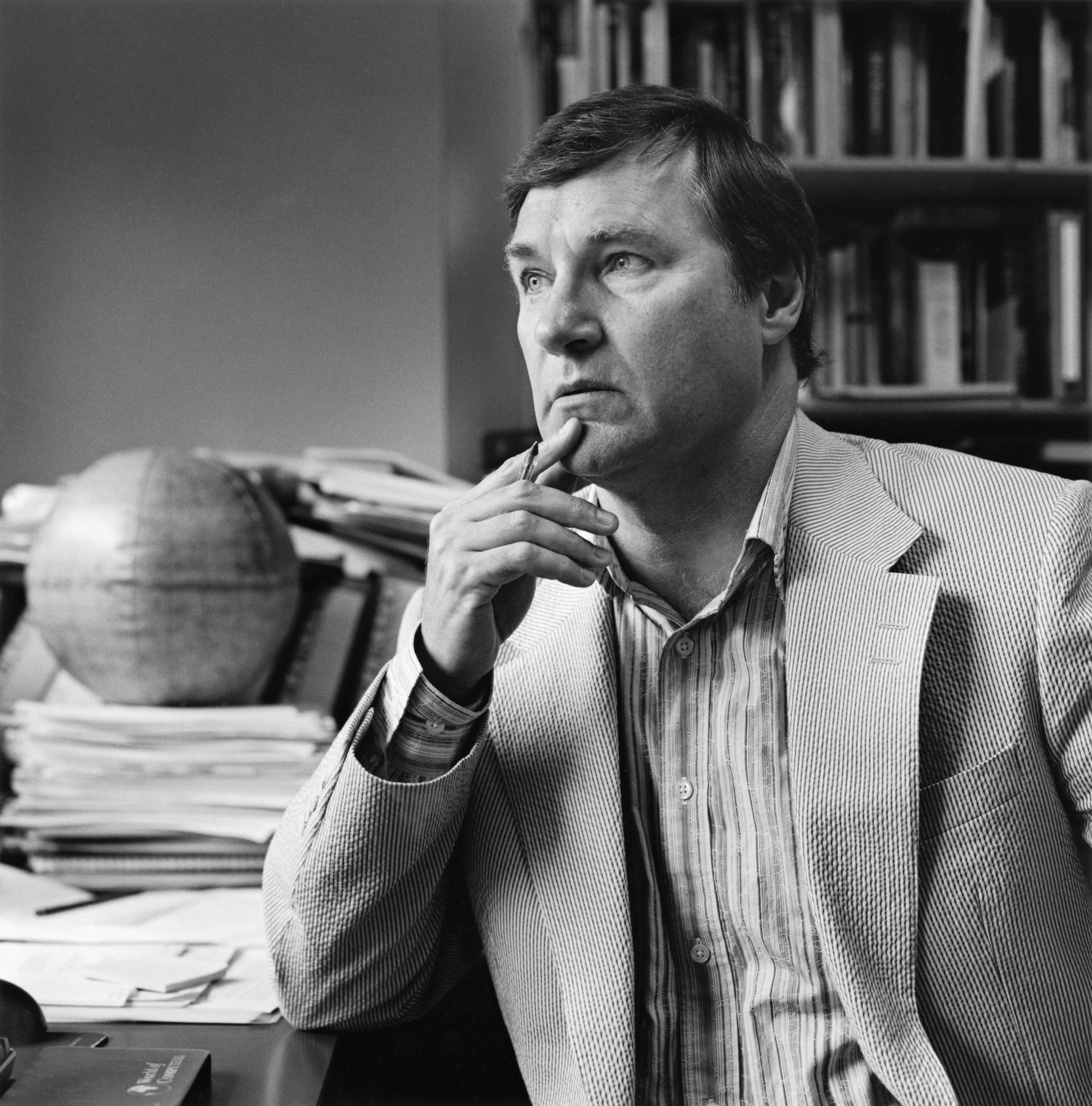}
    \caption{John David Barrow (born 1952), British cosmologist, theoretical physicist, and mathematician. Barrow was a Professor of Mathematical Sciences at the University of Cambridge since 1999. He was elected Fellow of the Royal Society in 2003 and was awarded the Faraday Prize in 2008 for his work on science popularisation. Photographed in 2007 in Barrow's office at the Centre for Mathematical Sciences, Cambridge. Photograph provided by Simon Stone, Science Photo Library, London.}
\end{figure*}

John David Barrow was born on 29 November 1952 to parents Walter and Lois. He had one sibling, an elder sister, Brenda, who pre-deceased John earlier this year. He grew up in Wembley, London, attending Barham Primary School and then Ealing Grammar School for Boys from 1964-71. Whilst his interest in science was sparked at the age of 12 when he was given a chemistry set, it must have seemed to the young John Barrow that a sporting career was far more likely. He was an outstanding runner, beating Steve Ovett at Withdean Stadium in Brighton in 1971 to become the English Schools 800m champion (over the same distance for which Ovett went on to win Gold at the 1980 Moscow Olympics). 
One of the authors, unaware of this precedent, was once flabbergasted to see the middle-aged John sprinting into the distance to catch a train. ``One never loses the fast-twitch muscles", he commented, laughing about this incident on his deathbed. But John's mind was even sharper than his muscles and in the end  prevailed in defining his career. \\ 

\begin{figure*}
    \centering
    \includegraphics[width=1.0\textwidth]{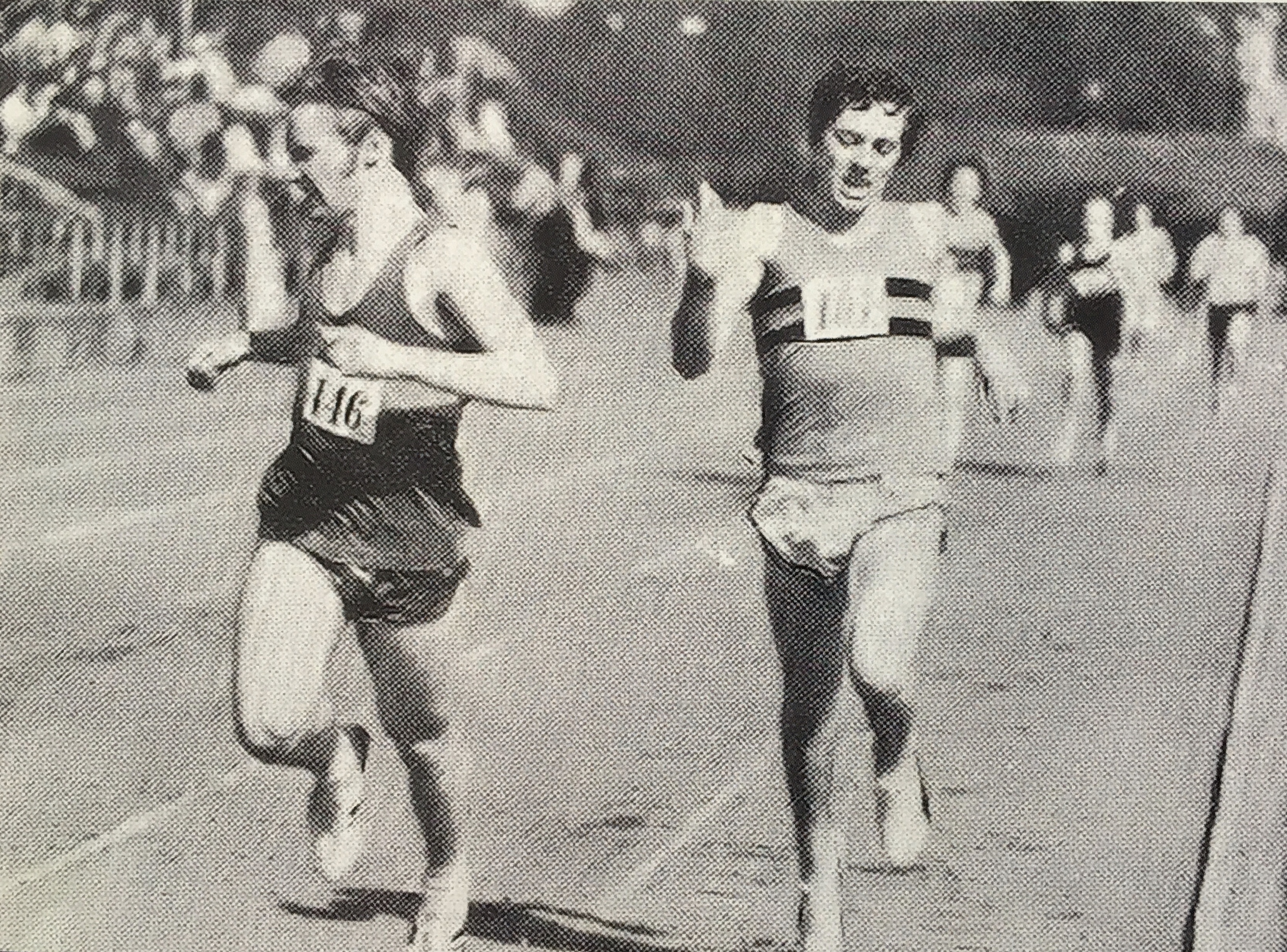}
    \caption{John beating Steve Ovett in the English Schools 800m, Withdean Stadium, Brighton, 1971.}
\end{figure*}

Even before finishing school, John had already met his future wife, Elizabeth. Like many of the baby boomer generation, John and his family were regular Sunday School attendees, where his teacher was one Mr. East, Elizabeth's father. The two children became close long before romance blossomed, although that had happened by the time university beckoned. Equipped with A-level Physics and Maths, John went up to Durham University to study Mathematics, whereas Elizabeth won a place at the Hammersmith Hospital Nursing School. There ensued regular train journeys over the next few years and when visits were on hold (at exam times John went into a self-imposed ``lockdown''), home-made food parcels were posted by Elizabeth, usually her delicious cakes. In 1974, John gained 1st Class Honours and was accepted at Magdalen College, Oxford, considering himself ``very lucky'' to be taken on as a graduate student by Dennis Sciama. This was the same year that Brandon Carter coined the term ``Anthropic Principle'', a highly controversial set of ideas that subsequently inspired John's remarkable 1986 book, co-authored by Frank Tipler, ``The Cosmological Anthropic Principle". As influential as Carter's paper was, it was clearly Barrow and Tipler's detailed study that garnered more attention for anthropic theories. But even though John was a pioneer in the field, he always showed a disarming lack of dogmatism about it. \\

John's brilliance had already emerged as a Durham undergraduate and it flourished at Oxford. He and Elizabeth were fortunate to find very convenient College accommodation and the couple married in 1975.  Their metamorphosis into a powerful partnership had begun, with Elizabeth ``carrying out the tasks that John could not and John doing what he did best'', as she put it herself.
In 1977 John was awarded his PhD, titled ``Non-Uniform Cosmological Models'', in the same year being awarded no less than Oxford's Wallace Research Prize, a Junior Research Lectureship at Christ Church College, Oxford, and a Lindemann Trust Fellowship. 
After brief negotiations with Christ Church, John and Elizabeth spent the first year of John's Lectureship in the Astronomy Department at the University of California Berkeley, returning to Oxford for the remaining 2 years. In 1979 he won (jointly with Bernard Carr) a Gravity Research Foundation award for the awesomely titled essay {\it ``Shear Hell Holes and Anisotropic Universes''}. John then returned to the Physics Department at Berkeley as a Miller Fellow for  a further year.\\

During these four years at  Berkeley and Oxford, John produced more than 30 papers, covering a bewilderingly diverse range of topics, including but not limited to, anisotropic cosmologies, black holes, big bang nucleosynthesis, and grand unification. The Berkeley years also saw the birth of Elizabeth and John's two sons, David and Roger. In 1981 John was offered a Lectureship in the Astronomy Centre at the University of Sussex. He remained at Sussex for the next 18 years, becoming a full professor in 1989. It was also there that John and Elizabeth completed their family, with the birth of their third child, Louise, in 1984.\\

The Sussex years saw John's writing career skyrocket, turning him into one of the world's foremost popularisers of science. His first book, {\it ``The Left Hand of Creation''} (in collaboration with Joe Silk) appeared in 1983.  John's seemingly endless energy resulted in 22 books over the next 37 years, covering the interplay between maths, cosmology, and the arts. In 1999, John's outstanding capacity for science communication attracted him to become the the Director of the Millennium Mathematics Project at Cambridge University, a Professor of Mathematical Physics, and a Fellow of Clare Hall. He remained at Cambridge for the rest of his life. Under John's guidance, the MMP project developed into a BBC collaboration, reaching over 1.5 million UK schoolchildren, parents and teachers. \\

\begin{figure*}
    \centering
    \includegraphics[width=1.0\textwidth]{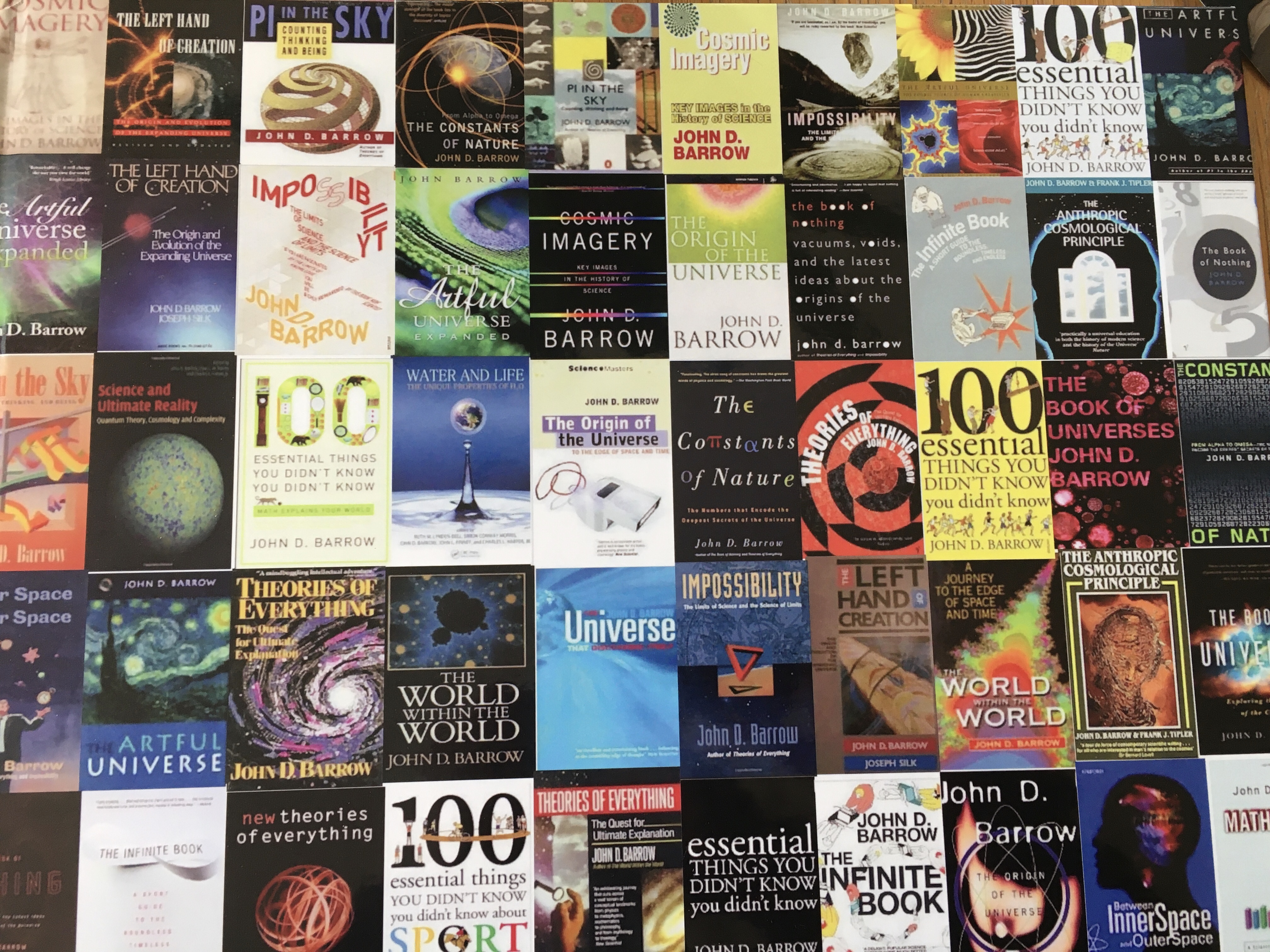}
    \caption{Some of John’s many books.}
\end{figure*}

John was made a Fellow of the Royal Society in 2003. He wrote over 550 scientific articles,  becoming an acknowledged world expert in topics as disparate as anisotropic models of the universe (the connection with chaos theory, for example), inflation (intermediate inflation models and the definition of slow-roll parameters), and the theory of cosmological singularities (the concept of sudden singularities). He was the undisputed world leader in theories of ``varying constants'', pioneered by Paul Dirac in the 1930s. In close relation to this, it is possible that John's greatest scientific passion was understanding how such a small number of natural laws and parameters could result in such complex outcomes: biology, consciousness, even nosy physicists laying claim to understanding the origin and evolution of the universe. He eagerly and objectively sought to know whether the laws of nature were programmed into the universe and immutable, or if instead the universe just ``made it up as it went along''. If he saw religious overtones in this quest, he never voiced it to either of us. John received the Templeton Prize in 2006 and in 2019 he was appointed to the Pontifical Academy of Sciences by Pope Francis. \\

But above all we will miss John Barrow, the person.  John had a unique sense of humour, sometimes mischievously so. He was very open-minded scientifically, combining a deceivingly conservative demeanour with an irrepressible gleam in the eye in the face of the wildest speculations. Everything was fair game to him. And although he was never dogmatic about his science, he was also firm in the face of irrational rejection. One of the authors keeps a trove of peer-review correspondence in which John sometimes reveals hilarious nonchalance and lack of emotion in response to the basest insults from irascible referees. 
He was the ultimate democrat. He treated students and senior scientists with equal respect and interest. In some ways he shone even more impressively in the company of non-scientists, because of his remarkable ability to converse logically at any level on almost any topic. He even lectured the British cabinet on chaos theory, at the request of Mrs. Thatcher.\\

We will also remember John as a bon-vivant. He was fond of haute cuisine and was a member of a chocolate tasting club (his sons complain that he used to hide the best ones). He was an enthusiast of music and theatre, with a definite artistic vein, so much so that in 2002 he penned his own theatre script {\it ``Infinities''}.
His play was performed that same year in Milan (in Italian) and  in Valencia (in Spanish), receiving multiple accolades (for example, the 2002 Italian Play of the Year Award).  He loved beautiful Italy and charming Italy loved him. It is telling that his last holiday, a few weeks before his demise, was in that  country. \\

We were privileged to know John Barrow, the polymath, not only as an admired professional collaborator, but also as a steadfast friend. Quite apart from intellectual excellence and creativity, few people combine generosity, courtesy, sincerity, and humour to the same degree. His memory retention was second to none and he was frequently referred to as a ``walking encyclopaedia''. One of us tried numerous times to entice him to one of the local village pub quizzes, albeit unfairly on the opposition, although he never did. John leaves behind his wife Elizabeth, to whom he was married for 45 years, as well as 3 children and 5 grand-children. On the day of his death, four year-old Poppy was seated at the family table in order for Elizabeth to explain that Grandpa had left forever and gone up to the stars. As the news was broken, she stared at the table top looking concerned and quickly said ``Oh no, Grandma! He's forgotten to take his mobile phone with him.''\\

With or without wireless coverage, John will always be with us. 

\end{document}